# Maintaining Constant Pulse-duration in Highly Dispersive Media using Nonlinear Potentials

*Haider Zia ***

Haider Zia
University of Twente, Department Science & Technology, Laser Physics and Nonlinear Optics Group, MESA+ Research Institute for Nanotechnology, Enschede 7500 AE, The Netherlands
E-mail: h.zia@utwente.nl

**Abstract**

A method is shown for preventing temporal broadening of ultrafast optical pulses in highly dispersive and fluctuating media for arbitrary signal-pulse profiles. Pulse pairs, consisting of a strong-field control-pulse and a weak-field signal-pulse co-propagate, whereby the specific profile of the strong-field pulse precisely compensates for dispersive phase in the weak pulse. A numerical example is presented in an optical system consisting of both resonant and gain dispersive effects. Here we show signal-pulses that do not temporally broaden across a vast propagation distance, even in the presence of dispersion that fluctuates several orders of magnitude and in sign (for example within a material resonance) across the pulse's bandwidth. Another numerical example is presented in normal dispersion telecom fiber where the length where an ultrafast pulse does not have significant temporal broadening is extended by at least a factor of 10. Our approach can be used in the design of dispersion less fiber links and navigating pulses in turbulent dispersive media. As well, we illustrate the potential for using cross-phase modulation to compensate for dispersive effects on a signal-pulse and fill the gap in the current understanding of this nonlinear phenomenon.

**Introduction**

The rich and complex physics of soliton formation has been a central topic in ultrafast optical physics since the early 1970s when Akira Hasegawa first suggested that temporal solitons could exist in anomalous optical fibers.[1,2] The initial allure of optical solitons was the prospect of a substantial improvement of data transmission rates in optical fiber communications since dispersive effects would not distort and smear out the solitary optical waveforms.

Optical solitons are also used in many bridging studies between the various fields of physics, such as optical analogs of bound states and Bose-Einstein condensates and even the simulation of black hole event horizons.[3–5] While higher-order solitons have wide use in nonlinear optics, such as in supercontinuum generation [6–9] and nonlinear pulse compression,[10–13] they are unstable and usually decay to fundamental solitons.[14] The fundamental soliton provides the most appeal, as its amplitude profile in the temporal domain does not change as it propagates.

However, the fundamental soliton is highly specific in waveform and is only stable under slow-varying group-velocity dispersion. I.e., the soliton profile must be a hyperbolic secant and its energy must satisfy specific material conditions.[14] The profile specificity presents limitations in potentially lucrative optical applications for solitons. For example, it would sometimes be more beneficial for optical waveforms that are not hyperbolic secant profiles, e.g., Gaussian pulses, to propagate in dispersive media without temporally broadening.

Recently, this problem has been somewhat resolved as soliton solutions in the presence of even higher-order dispersion terms have been analytically found and experimentally demonstrated.[15] Nevertheless, while the choice of pulse profiles is now larger, it can still only be one of specific profiles. As well, to exploit these higher-order dispersion solitons, other dispersion terms cannot be present past a perturbative degree. For example, the so-called quartic soliton must only be accompanied by a nonzero 4th order group-velocity dispersion (GVD) coefficient.

Given the above discussion, it would be highly beneficial to obtain a method to generate stationary or close to stationary pulse profiles (i.e., where the shape of the pulse and its duration remains approximately the same for the entire propagation) for a more significant amount of pulse shapes and dispersion profiles. The technique would enable, for example, pulses that could be guided through turbulent group-velocity dispersion that fluctuates both in sign and in orders of magnitude without broadening. Examples of turbulent dispersive media are widespread in optics and they include resonant loss and gain media, Raman active materials and past the photonic bandgap of Bragg grating structures. [16–20] Another application could be to prevent temporal broadening of complicated single-photon wavefunction shapes (that are far from hyperbolic-secants) such that dispersive effects are negligible in resonator configurations used as quantum memory. [21,22]

In this paper, we formulate and show, through numerical examples, a method to prevent temporal broadening of pulses of more arbitrary shapes in a manner less dependent on the waveguide's dispersion profile than current soliton generation allows. We do this by controlling a weak-amplitude signal-pulse with a 'control' pulse that offers an optical potential that induces cross-phase modulation (XPM).

XPM has been explored before in the context of solitons, for example, in generating (second-order dispersion) soliton pairs across dispersion regions, i.e., a dark soliton in normal dispersion and a bright soliton in anomalous dispersion that is sustained through XPM. Cross-phase modulation has also been extensively explored in the context of continuously shifting the wavelength of a pulse or a dispersive narrowband spectral component in a technique called optical trapping. [23–28]

While these formative experiments are significant to study XPM and enable new applications, they do not overcome the pulse profile specificity needed. For either the control or signal-pulse or both a highly specific profile, such as a fundamental soliton profile, or a truncated Airy function[29] must be present. Also, these approaches only work with dominant second-order group-velocity dispersion, where other dispersive terms are negligible.

As opposed to the previous literature that primarily focused on the frequency effects of XPM or interaction dynamics of specific pulse shapes, our focus here is to use XPM to limit the temporal broadening of a general signal-pulse in a general dispersion profile. The analytic expression for the exact control-pulse profile needed given an arbitrary or general signal-pulse profile is obtained for the first time. The control-pulse shape is obtained that would generate the conjugate temporal phase to the one from dispersion, such that the two cancels to zero for the signal-pulse.

Ultimately this allows for near invariant temporal solutions across a wide range of signal-pulse shapes and dispersion profiles. The method then alleviates the burden of trial-and-error methods to find the proper control-pulse profile and fills the gap in our understanding of XPM induced control of temporal broadening.

## 1. Theoretical Description

We start by deriving the method's expressions used to obtain the control-pulse needed for maintaining a stationary signal-pulse. We restrict our analysis to the waveguide case, where only the propagation spatial dimension needs to be considered. The nonlinear equations of motion of two pulses, separated in central frequency or polarization, with complex envelope functions $u_1$ (control-pulse) and $u_2$ (signal-pulse) undergoing XPM is given in **Equations** (1) and (2):[30]

$$\frac{\partial u_1}{\partial z} = \sum_{k \geq 2} \frac{i^{k+1}}{k!} \beta_k \frac{\partial^k}{\partial T^k} u_1 + i\gamma[|u_1|^2 + 2|u_2|^2]u_1$$

$$\frac{\partial u_1}{\partial z} \approx \sum_{k \geq 2} \frac{i^{k+1}}{k!} \beta_k \frac{\partial^k}{\partial T^k} u_1 + i\gamma[|u_1|^2]u_1$$

(1)

$$\frac{\partial u_2}{\partial z} = \sum_{k \geq 2} \frac{i^{k+1}}{k!} \kappa_k \frac{\partial^k}{\partial T^k} u_2 + i\gamma[|u_2|^2 + 2|u_1|^2]u_2$$

$$\frac{\partial u_2}{\partial z} \approx \sum_{k \geq 2} \frac{i^{k+1}}{k!} \kappa_k \frac{\partial^k}{\partial T^k} u_2 + i\gamma[2|u_1|^2]u_2$$

(2)

we impose that $|u_1|$ is much larger than $|u_2|$ and that the signal-pulse has a small amplitude such that it does not contribute to any nonlinear effects. Therefore, Equations (1) and (2) reduce to the indicated approximate forms. $\beta_k$, $\kappa_k$ are the dispersive coefficients, arising from the frequency-dependent wavenumber Taylor expanded about the control or signal's central frequency $\omega_1$ or $\omega_2$. $T$ is the pulse time-coordinate defined in a co-moving frame-of-reference.

The above system-equations are solved to obtain the control-pulse envelope profile giving a stationary solution of the signal-pulse. However, a few conditions must be met, which we list before indicating the derivation:

1. The dispersive broadening of the control-pulse must be negligible, i.e., the dispersion length, usually approx. proportional to $\frac{1}{\beta_2}$ [6], must be larger than the interaction length of the two pulses.

2. The group-velocities of both pulses, $u_1$ and $u_2$ are the same. This means that the group-delay at the control-pulse's central frequency, i.e., $\frac{\partial k}{\partial \omega}|_{\omega_1}$ where $k$ is the wavenumber, must equal to that of $u_2$.

The ideal group-delays of the pulse-pair then satisfy,

$$\int_{\omega_0}^{\omega_1} \tilde{D}_1 d\omega = \int_{\omega_0}^{\omega_2} \tilde{D}_2 d\omega \quad . \qquad (3)$$

Where $\tilde{D}_1$, $\tilde{D}_2$ are the frequency dependent group-velocity dispersion seen by the control and signal-pulse in **Equation** (3). $\omega_0$ is the frequency corresponding to the group-velocity of the common moving time frame-of-reference of the two pulses. We take a frame-of-reference co-moving and centered with the control-pulse (i.e., $\omega_0$ is the control-pulse's central frequency).

The group-velocity , given by Equation (3) is based on the integral of the group-velocity dispersion; therefore, many dispersion profile arrangements could satisfy the condition for the group-velocities of the two pulses to be equal. Then matching the group-velocity of the pump and signal is not constrained to a highly specific waveguide. The non-constrained nature of GV-matching is illustrated by the large amount of existent literature where nonlinear effects such as second harmonic generation and cross-phase modulation are observed by the interaction of GV-matched pulses.[31],[32],[33],[34],[35] In fact, for radially symmetric waveguides (such as fiber) a simple trick to insure the same group-velocity is to set the pump and signal pulse to orthogonal polarizations, as done in section 3.

Likewise, the condition that dispersive effects of the control pulse are negligible - in comparison to those seen by the signal pulse – is satisfied by a wide variety of waveguides explored in past studies. For example at the zero-dispersion point (or region) of photonic crystal fibers or at a wavelength range in the detuned region of a Bragg grating structure.[36],[37],[38],[39] This would then be the location of the control pulse. However, for other wavelengths, the dispersion profile can oscillate by orders of magnitude specially around resonances of the waveguide system (for example, see [40] or within the bandgap of a Bragg-grating structure). Thus, the oscillating part of the dispersion profile would be where the signal pulse is located (as a corollary, the signal pulse and control pulse would usually then have central wavelengths separated by a large range). Such general systems, where the dispersion is turbulent for a signal pulse but well-behaved for a control pulse are the discussion point of section 2.

1.1. Control-pulse Formulation

The first step in formulating the required control-pulse is to insert the following Ansatz $u_2(T,z) = u_2(T, z=0)e^{ipz}$ ( $p$ is a constant) into Equation (2). The approximation : $u_1(T,z) = u_1(T, z=0)e^{i\gamma |u_1|^2 z}$ is also inserted –which is exact if there is no dispersive broadening for the control-pulse. Dividing out common terms, this yields

$$|u_1(T, z=0)|^2 = \frac{1}{2\gamma}\left[ p - \frac{1}{u_2(T, z=0)} \sum_{k \geq 2} \frac{i^k}{k!} \kappa_k \frac{\partial^k}{\partial \tau^k} u_2(T, z=0) \right] \qquad (4)$$

Since the left-hand side of **Equation** (4) equates to a positive definite function for all $T$, so must the right-hand side. From this positive definitive condition, $p$ is a real constant and it can be shown that the signal-pulse must always be in a form $u_2(T, z=0) = f(T)e^{i\nu T}$ where $f(T)$ is a transform limited complex amplitude function and $\nu$ is a constant. Generally,

$$p \geq \max \left[ \frac{1}{u_2(T,z=0)} \times \sum_{k \geq 2} \frac{i^k}{k!} \kappa_k \frac{\partial^k}{\partial \tau^k} u_2(T,z=0) \right],$$ to guarantee that the left-hand side can never be negative.

It seems mathematically, that another condition needed for Equation (4) to be valid is that the odd-dispersion coefficients must be zero, at least, for the dispersion profile that covers the signal-pulse's bandwidth. Thus, across the bandwidth of the signal-pulse, the dispersion profile is ideally symmetric about the central wavelength – e.g, for ultrashort <100 fs pulses, the dispersion profile should be symmetric across a bandwidth range less than 50 nm. There are a plethora of waveguides that satisfy this symmetry requirement from dispersion engineered photonic crystal fibers to the system described in [15]. Furthermore, we find from full numerical simulations that the symmetry condition is not strictly necessary (see section 3 for related discussion) and introduces little error for moderate odd-order dispersion coefficients (taken about the central wavelength of the signal) if Equation (4) is modified to Equation (5) given as:

$$|u_1(T,z=0)|^2 = \left| \sqrt{\frac{1}{2\gamma} \left[ p - \frac{1}{u_2(T,z=0)} \sum_{k \geq 2} \frac{i^k}{k!} \kappa_k \frac{\partial^k}{\partial \tau^k} u_2(T,z=0) \right]} \right|^2 , \qquad (5)$$

even if this modification is not supported in a strict mathematical sense.

Equation (4) indicates that $u_1(T,z=0)$ can be a product between any real function and exponential phase function since only the absolute magnitude must obey the equation. Values of $T$ that lie outside of the temporal profile of the signal-pulse, $u_1(T,z=0)$ can equal zero or any convenient function without additional error.

1.2. Deviation From Stationary Propagation

In a practical setting, since Equation (4) consists of a quotient term, it may be necessary to truncate it for large $T$ when $u_1(T, z = 0)$ goes to zero. The truncation is decided by the desired energy upper bound that the control-pulse must satisfy. Thus, $u_1(T, z = 0)$ is uncontrolled for $T$ larger than the truncation window, yielding the potential error of dispersive broadening in the wings of the signal-pulse.

For example, it can be shown that for a Gaussian pulse, $u_1(T, z = 0) = Ae^{-\frac{T^2}{2T_o^2}}$, Equation (4) becomes

$$|u_1(T, z = 0)|^2 = \frac{1}{2\gamma}\left[p - \sum_{k \geq 2} \frac{i^k}{k!} \kappa_k H_{e_k}\right], \tag{6}$$

where $H_{e_k}$ are the set of Physicists Hermite Polynomials[41] of order $k$.

When only second-order group-velocity dispersion is present, **Equation** (6) equates to a parabolic function, and therefore, does not converge uniformly to zero. Thus, the edges of a Gaussian pulse may temporally disperse away from the XPM controlled central region of the signal-pulse. This limitation will be explored in the subsequent section.

## 2. Near Stationary Pulses in Turbulent Dispersive Media

To understand both the limitations and impact of the above theoretical procedure for XPM induced stationary pulse propagation, we consider pulse propagation in a turbulent dispersive waveguide medium. We restrict our analysis to the waveguide case, as diffraction is negligible and can be omitted from the analysis. The diffractive term of the full three-dimensional NLSE, for example shown in [42] cannot be accounted for by method used to obtain Equation (4).

To solve for the dynamics of the pulses undergoing nonlinear propagation in the turbulent waveguide, we first use Equation (5) to generate the amplitude function of the control-pulse given a certain signal-pulse we want to propagate invariantly. We then numerically solve, using the split-step Fourier method, [30] the coupled NLSE equations (Equations (1) and (2)) for both pulses, with the input control and signal profile as initial conditions.

The turbulent dispersive waveguide medium we study is defined as having narrow fluctuations between positive and negative GVD spanning orders of magnitude within a small bandwidth of frequencies. We apply our method for turbulent dispersion because the highly fluctuating dispersion demonstrates the full effectiveness and potential for our method, i.e., the large dispersion fluctuations would make controlling the associated dispersive effects challenging. While we show our method in this extreme case, we find our method is effective for more moderate examples too, such as pulse propagation through a telecom fiber with modest higher-order dispersion.

Impactful examples of such highly fluctuating dispersion that occur in many optical applications are generally when the frequency range of a pulse covers a waveguide's absorption dip and/or gain peak or covers waveguide resonances, e.g.., from leaky modes or evanescent coupling to neighboring waveguides. Such a case can occur for an ultrafast pulse (i.e., covering a large bandwidth) traveling through a doped waveguide that amplifies light while also coupled to a ring resonator detuned from the gain profile.[17] Another significant case is when a pulse travels in a waveguide where the Stokes peak location and absorption dip location (Stokes-shift) from the Raman effect are situated within the bandwidth of the original pulse.[19]

To illustrate the control of temporal broadening within turbulent dispersive media, we have constructed an artificial waveguide that possesses a weak absorption and gain resonance characteristic of leaky mode strand resonances in photonic crystals. [40] The features are characterized by a Lorentzian gain peak and absorption dip within the bandwidth of an ultrafast (200 fs) pulse, where the pulse has a carrier frequency that sits precisely at the midpoint of these features. We take as the Lorentzian parameters used, the typical values from leaky mode strand resonances in photonic crystal fibers given in [40]. While we show an example with only two resonant features, we have also studied waveguides consisting of a series of multiple resonances, whereby the pulse bandwidth is centered such that the dispersion profile is symmetric within its bandwidth.

The waveguide has a flat-zero GVD when the absorption and gain are neglected. Through the associated Kramers-Kronig relations, the group-velocity dispersion is derived in the waveguide and is shown in **Figure 1a** along with the signal and control's normalized spectral energy density distributions. The dispersion is symmetric around the signal-pulse's carrier frequency and fluctuates over five orders of magnitude.

| a) | b) |
|---|---|
| 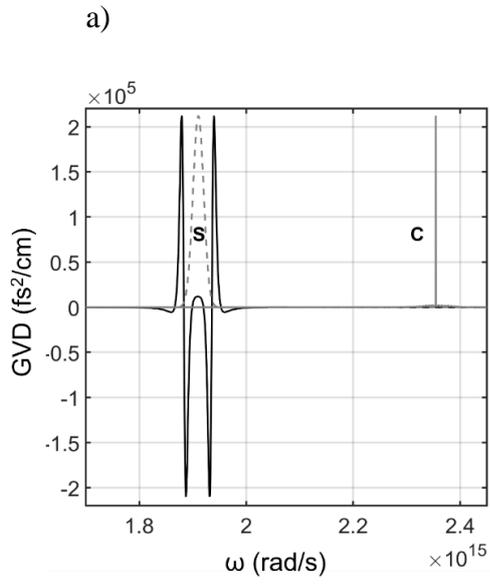 | 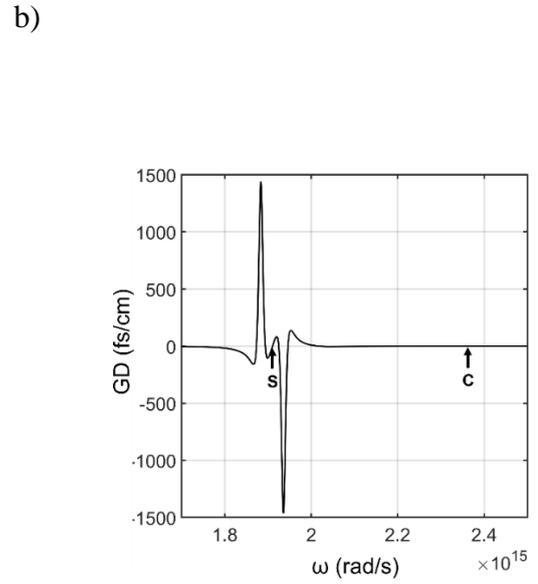 |
| c) | d) |
| 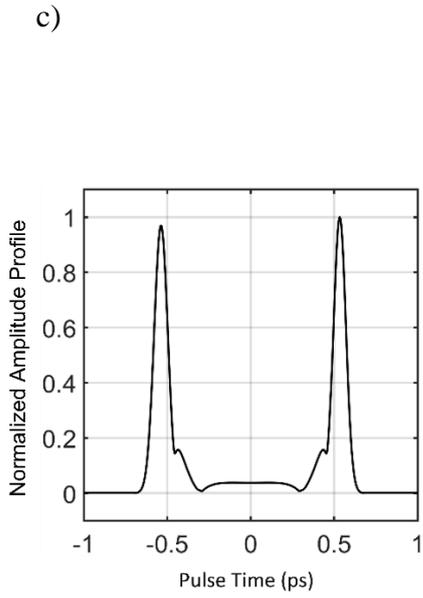 | 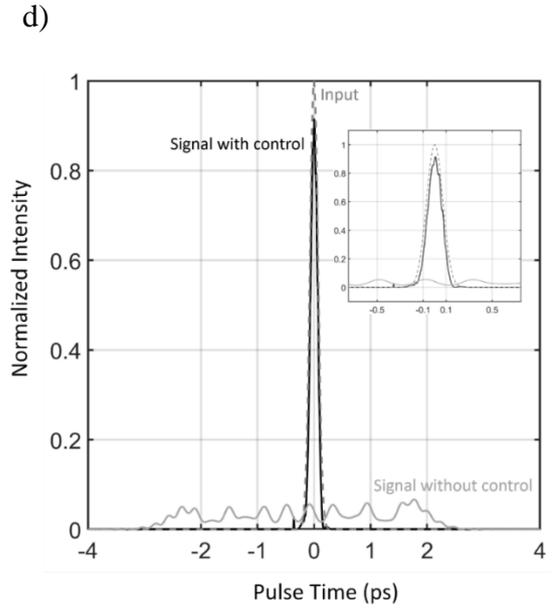 |
| e) | f) |
| 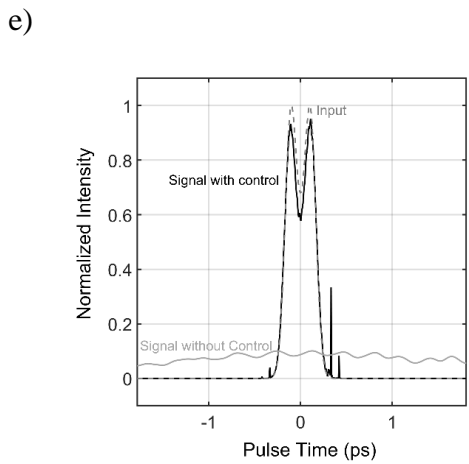 | 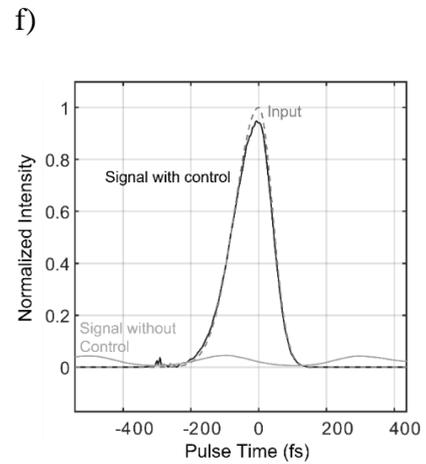 |

**Figure 1.** a) Group-velocity dispersion versus angular frequency with signal (s) and control (c) normalized spectral-density-distributions indicated in gray. b) Frequency-dependent group-delay with signal and control-pulse locations indicated. c) Amplitude of control-pulse versus pulse time. d) Normalized-to input-peak intensity distribution versus pulse-time for signal without control, signal with control and input profile. Inset is a zoom-in around the controlled signal-pulse.

The interaction of the highly fluctuating dispersion with our set of signal-pulses provides a stringent test for how well the XPM modulated signal-pulse avoids temporal broadening. To evaluate the quality of the controlled signal-pulse we compare it to the signal's temporal profile when there is no control-pulse. This comparison is done for three different signal-pulse types to show that our method works for arbitrary signal-pulses.

To isolate the dynamics of the signal-pulse, we center the control-pulse in a zero-dispersion region such that dispersive distortion of the amplitude profile is not present. As well, both signal and pulse propagate at approx. the same group-velocity with negligible walk-off as shown in the group-delay versus frequency, i.e., **Figure 1b**. The amplitude profile of the control-pulse is shown in **Figure 1c**, with an energy of approx. 11 µJ. In general, we find that the control-pulse energy scales directly with the order of magnitude of the dispersion fluctuations, going to below nJ's for dispersion fluctuations on the order of unity.

We now proceed with our first signal-pulse example, which is a 200 fs Gaussian pulse. After 25 cm of propagation in the medium, the XPM controlled signal-pulse remains close to the Gaussian input pulse in profile and duration as shown in **Figure 1d**. For comparison, the signal without the control-pulse is also shown, where it temporally broadens to a window of 6 ps and is highly distorted from the original Gaussian profile.

The next signal-pulse example is far from the typical Gaussian shape as shown in **Figure 1e**. However, as in the Gaussian pulse case, the pulse maintained its duration and shape close to the input profile's parameters. Without the control-pulse the pulse would be highly distorted in shape, stretching out to 5 ps.

We then proceed with the last example which is an asymmetric pulse that is slanted towards positive pulse time, as shown in **Figure 1f**. As with the previous cases, the XPM controlled signal-pulse is close to the input profile at the end of propagation, while the uncontrolled case distorts considerably and stretches to approx. 6 ps.

Interestingly, the effects of the uncontrolled wings of the pulses (see section 1.1 for associated discussion) can be seen as a slightly reduced pulse duration from the input pulse and small amplitude pre/post pulses (below the -4 dB level). This type of error then contributes to primarily to the outcome of pulse compression instead of broadening, desirable for many applications. The dynamics at the temporal edge regions of the control-pulse explains the emergence of these deviations.

The edge locations of the control-pulse are defined when the input signal-pulse goes below an absolute amplitude value of -50 dB. The so-defined decaying edges of the control-pulse introduces a chirp profile that compresses the dispersively chirped temporally broadening wings of the signal-pulse and can even break them into sub-pulses as they interact with it, causing the overall pulse compression.

Also, there is a slight mismatched group-velocity of the signal with control-pulse because of numerical floating-point error. Though small, the mismatch produces XPM induced wavebreaking [43] that adds small modulations on one side of the pulse profiles. This numerical error also enhances the error from the edge effects.

We provide a video of the second example pulse's propagation across the waveguide in supplementary. The video illustrates the maintained duration and shape of the signal-pulse with control. The specific control pulses for the signal pulses shown in Figure 2 and the associated p-values are given in Supplementary Information 2, in **Table S1** and **Figure S2a,b**.

## 3. Impact of Asymmetric Dispersion on Signal and Dispersive Effects on Control Pulse

The previous section shows how the method can control pulses in highly dispersive media. However, a noteworthy trait of the illustrated waveguide GVD in section 2 is that it is symmetric about the signal pulse's bandwidth, as this is a requirement for Equation (4). This section will explore a modification of Equation (4), given as Equation (5), to account for cases where there is an asymmetric dispersion profile for the signal pulse.

To illustrate how our method can work when there is moderate odd-order dispersion, we take the GVD profile and the nonlinear coefficient of a well-known radially symmetric step-index fiber found in many telecom applications. [11] The dispersion profile of the fiber so-named "Corning Hi1060Flex" is given in **Figure 2a**. As can be seen in Figure 2a, the fiber has normal dispersion for the frequency range considered and there is a large asymmetry in the GVD profile of the fiber, about the location of the signal pulse central wavelength. In this study, we set the control and signal pulse to the same wavelength but orthogonal polarizations, so that they can still interact through cross-phase modulation. Since both signal and pulse see the same dispersion profile, the group-velocities are matched.

3.1 Asymmetric Dispersion Across Signal Pulse

We start by propagating a signal and control pulse, dictated by Equation (5), across 30 meters of the fiber waveguide. The signal pulse bandwidth is in a dispersion region where the dispersion profile is asymmetric about the pulse's central wavelength, and thus possesses moderate odd-order dispersion coefficients when Taylor expanded around this central wavelength. For this section, the control pulse's dispersive effects are turned off to properly illustrate the dynamics of a signal pulse centered in a non-symmetric dispersion profile range.

**Figures 2b-c** show that the control pulse successfully prevented the signal pulse from dispersing, for a variety of signal pulse shapes (Gaussian, two-peak pulse), to show the general

validity of the method. For comparison the signal without control is shown in Figure 2b, indicating large dispersive effects with no control pulse present.

3.2 Asymmetric Dispersion Across Signal Pulse and Significant Presence of Dispersive Effects for Control pulse

When the control pulse also undergoes significant dispersive effects, i.e., comparable or equivalent to what the signal sees, the ability for it to prevent dispersive broadening for the signal is lowered. In this fiber case, the control pulse sees the same GVD as that of the signal pulse, thus, dispersive effects for the control pulse are significant. **Figure 2d** indicates the results of the propagation of control and signal over 10 meters of fiber and compares to the case where the control pulse is not present. It is found that under the effects of dispersion, the control pulse can still prevent significant signal pulse temporal broadening up to 10 meters. At 10 meters of propagation, the signal pulse broadens to a factor of 1.8, versus a factor of 19 of temporal broadening when the signal propagates without control. Therefore, the dispersion length of the signal is increased by approximately a factor of 10 when the control pulse is present. The control pulse used for this simulation is the same as the one displayed in Figure 2b.

While temporal broadening is prevented, the signal pulse suffers from modulations on its amplitude profile due to distortions of the control pulse's amplitude profile from dispersion. The control pulse has a dip to zero at the center as shown in Figure 2b. Under dispersion and self-phase modulation, the two peaks broaden into the dip, resulting in the dip depth being lowered and the slop of the dip being increased (until approx. 1 meter of propagation) and then decreased. The dip dynamics of the control pulse then has the result that the signal pulse temporally narrows in duration from the input duration, which is advantageous for certain applications, and then slowly broadens to the duration seen in Figure 2d. As well, because of the high third order dispersion, the control pulse undergoes an asymmetric broadening of its profile; resulting in the dip becoming asymmetric at the center. The asymmetry in the control pulse results in Airy like side-lobs in the signal pulse, as well as a global shift in central wavelength of the signal pulse through cross-phase modulation.

The resultant effect of the wavelength shift, results in a global delay of the signal pulse with control versus the signal without control. We shifted the signal with control pulse back to time zero in relation to the moving frame of reference of the simulation, so that the comparison with input and dispersed signal is clearly visible in Figure 2d. See Supplementary Section 1 Figure S1, for the view of the signal with control at 1 meter of propagation, where narrowing of the duration is visible, along with Airy side-lobs and the delay of the signal pulse with control.

The overall energy of the control pulse was approx. 800 pJ. Please see Supporting Information 1 for further discussion on further modifications for the control pulse under this dispersive case, the control pulse is given in **Figure S1**. The specific control pulses for the signal pulses shown in Figure 2 and the associated p-values are given in Supplementary Information 2, in **Table S1** and **Figure S2**c. By comparing values of the fiber case with the turbulent dispersion medium of section 2, it is found that the general trend for the p-values is that they scale proportionally to the maximum GVD magnitude within the frequency range of interest.

The results of the numerical experiment illustrated in Figure 2d, indicates that an ultrafast pulse can be prevented from substantial temporal broadening up to meters of length in significant normally dispersive fiber. The results are based on the parameters of a popular telecom fiber for 1550nm radiation used in telecommunications. A practical application is for example that, dispersive broadening which would limit fiber links between, different labs at a university can now be utilized with the presence of a control pulse using the method in this paper.

a)

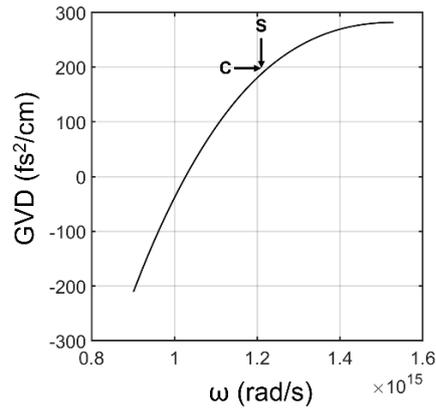

b)

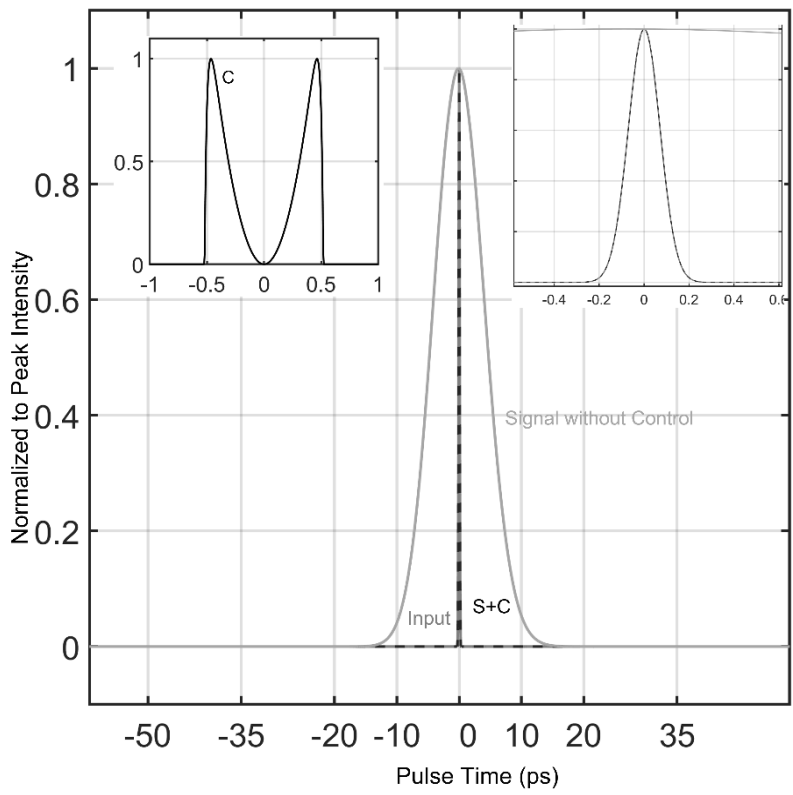

c)

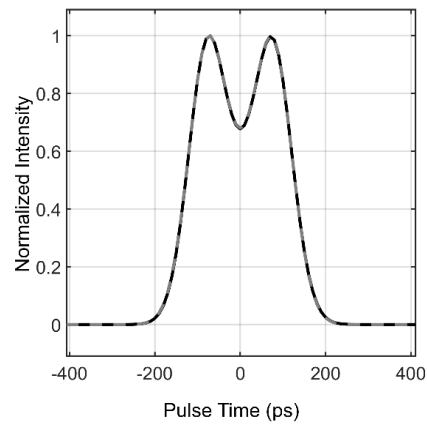

d)

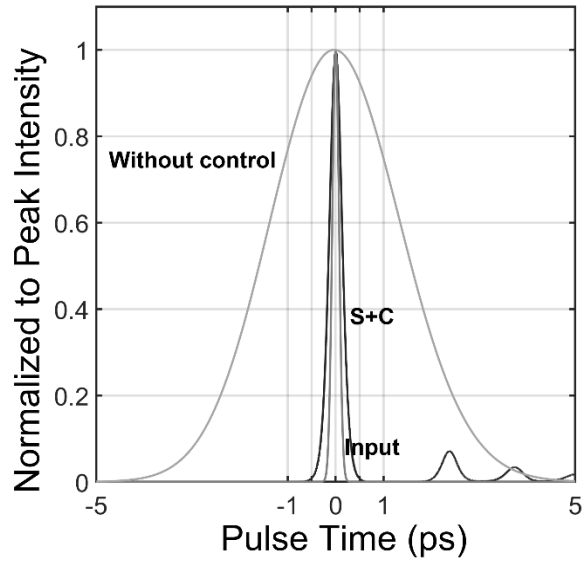

**Figure 2:** a) The group-velocity dispersion versus angular frequency. The frequency of the control (c) and signal (s) is marked by a vertical and horizontal arrow to indicate the respective polarization eigenstates. b) The normalized to peak intensity of the signal in the presence of control pulse (s+c), the input profile, and the signal without control pulse at the end of a 30 meter fiber. The left inset indicates the control pulse's normalized intensity; the right inset is a zoom-in of the input and signal. c) Normalized to peak intensity for two pulse shapes chosen ad hoc (signal in black, input profile in dotted gray). d) Normalized to peak intensity of the 200 fs Gaussian signal pulse propagated across 10 meters of the fiber compared to the input profile and the case without control. Here, the control pulse has full dispersive effects.

**Outlook**

This paper so far obtains a method to obtain a control pulse to minimize dispersive effects in a signal pulse through cross phase modulation; however, practically obtaining said control pulse is not considered up to this point. We then dedicate a portion of the outlook to discuss the generation of the control pulse.

A popular avenue for ultrafast pulse shaping is to do it in the frequency domain via dynamic diffraction gratings that can modulate the amplitude (and phase ) of the spectral profile. These gratings are usually generated through an Acousto-optic modulation (AOM) effect found by sending radio-frequency electromagnetic radiation through a piezo-electric medium, surrounding an optically active crystal. The dynamic control of the RF signal induces a desired standing acoustic wave pattern in the crystal which then acts as a diffractive grating for the optical signal. The scheme ultimately enables the spectral profile of picoseconds down to tens of femtoseconds pulses to be shaped arbitrarily. Extensive literature has explored this subject and there are even products available to shape ultrafast pulses (with spectral bandwidths even exceeding 50 nm). [44],[45],[46] Then, this example scheme should be sufficient to generate the types of control pulses seen in this paper from a seed pulse (e.g, from a Ti:Sa mode-locked laser or an Erbium doped fiber laser).

Other interesting applications emerge if the amplitude of the control-pulse, whose shape is determined by Equation (4) is increased. Instead of just compensating for temporal broadening, the control-pulse acts to nonlinearly compress the signal-pulse, even in dispersive media where temporal self-compression, for example, due to self-phase modulation, is unlikely, i.e., in normal dispersion.

Our calculations indicate that XPM on a signal-pulse can enhance the spectral broadening (e.g., in normal dispersion waveguides) by a factor of $\sqrt{2}$. The temporal compression factor (e.g. in anomalous dispersion waveguides) is also enhanced by the same factor – simply due to this factor of two in the XPM nonlinear coefficient. Thus, using the control-pulse to induce spectral generation through XPM is potentially a highly efficient nonlinear process.

We preliminarily find that simply increasing the amplitude of the found control-pulse from the calculated value given by Equation (4), can induce nonlinear compression of the signal-pulse through the induced additive phase from cross-phase modulation. Thus, the presented method can be used to calculate ideal profiles of control-pulses for XPM on signal-pulses, for the control of such effects as temporal broadening. Also, furthermore, the here presented method could enhance nonlinear bandwidth generation and nonlinear pulse compression in weak amplitude signals.

**References**

[1] A. Hasegawa, and F. Tappert, *Appl. Phys. Lett.* **23**, 171–172 (1973).
[2] D. Grahelj, *Seminar* (2010).
[3] A. Demircan, S. Amiranashvili, and G. Steinmeyer, *Phys. Rev. Lett.* **106**, 1–4 (2011).
[4] A. V. Yulin, and D. V. Skryabin, *Phys. Rev. A - At. Mol. Opt. Phys.* **67**, 8 (2003).
[5] Y. V Kartashov, and V. A. Vysloukh, *Phys. Rev. E* **72**, 26606 (2005).
[6] K. Luke, Y. Okawachi, M. R. E. Lamont, A. L. Gaeta, and M. Lipson, *Opt. Lett.* **40**, 4823 (2015).
[7] K.-J. Porcel, Marco A.G., Scheppers, Florian, Epping, Joern, Hellwig, Tim, Hoekman, Marcel, Heideman, Rene G., Van der Slot, Peter J. M., Lee, Chris J., Schmidt, Robert, Bratschitsch, Rudolf, Fallnich, Carsten, Boller, *Opt. Express* **25**, (2017).
[8] J. P. Epping, T. Hellwig, M. Hoekman, R. Mateman, A. Leinse, R. G. Heideman, A. van Rees, P. J. M. van der Slot, C. J. Lee, C. Fallnich, and K.-J. Boller, *Opt. Express* **23**, 19596 (2015).
[9] H. Zia, N. M. Lüpken, T. Hellwig, C. Fallnich, and K. J. Boller, *Laser Photonics Rev.* **14**, 1–11 (2020).
[10] J. C. Travers, T. F. Grigorova, C. Brahms, and F. Belli, *Nat. Photonics* **13**, 547–554 (2019).
[11] H. Zia, *Photonics* **8**, 1–18 (2021).
[12] J. W. Choi, B. U. Sohn, G. F. R. Chen, D. K. T. Ng, and D. T. H. Tan, *APL Photonics* **4**, (2019).
[13] H. Chen, Z. Haider, J. Lim, S. Xu, Z. Yang, F. X. Kärtner, and G. Chang, **38**, 4927–4930 (2013).
[14] J. M. Dudley, G. Genty, and S. Coen, *Rev. Mod. Phys.* **78**, 1135–1184 (2006).
[15] A. F. J. Runge, Y. L. Qiang, T. J. Alexander, M. Z. Rafat, D. D. Hudson, A. Blanco-Redondo, and C. M. de Sterke, *Phys. Rev. Res.* **3**, 1–8 (2021).
[16] J. E. Heebner, and R. W. Boyd, *J. Mod. Opt.* **49**, 2629–2636 (2002).
[17] A. M. Steinberg, and R. Y. Chiao, *Phys. Rev. A* **49**, 2071–2075 (1994).
[18] P. Sprangle, J. R. Peñano, and B. Hafizi, *Phys. Rev. E - Stat. Physics, Plasmas, Fluids, Relat. Interdiscip. Top.* **64**, 5 (2001).
[19] J. E. Sharping, Y. Okawachi, and A. L. Gaeta, *Opt. Express* **13**, 6092 (2005).
[20] B. J. Eggleton, C. M. de Sterke, and R. E. Slusher, *J. Opt. Soc. Am. B* **14**, 2980–2993 (1997).
[21] T. B. Pittman, and J. D. Franson, *Phys. Rev. A - At. Mol. Opt. Phys.* **66**, 4 (2002).
[22] P. M. Leung, and T. C. Ralph, *Phys. Rev. A - At. Mol. Opt. Phys.* **74**, 1–6 (2006).
[23] M. N. Islam, G. Sucha, I. Bar-Joseph, M. Wegener, J. P. Gordon, and D. S. Chemla, *J. Opt. Soc. Am. B* **6**, 1149 (2008).
[24] S. Trillo, S. Wabnitz, E. M. Wright, and G. I. Stegeman, *Opt. Lett.* **13**, 871–873 (1988).
[25] N. Nishizawa, and T. Goto, *Opt. Express* **10**, 1151 (2002).
[26] S. F. Wang, A. Mussot, M. Conforti, X. L. Zeng, and A. Kudlinski, *Opt. Lett.* **40**, 3320 (2015).
[27] Z. Deng, J. Liu, X. Huang, C. Zhao, and X. Wang, *Opt. Express* **25**, 28556 (2017).
[28] S. F. Wang, A. Mussot, M. Conforti, A. Bendahmane, X. L. Zeng, and A. Kudlinski, *Phys. Rev. A - At. Mol. Opt. Phys.* **92**, 1–6 (2015).

[29] M. Goutsoulas, V. Paltoglou, and N. K. Efremidis, *J. Opt. (United Kingdom)* **19**, (2017).
[30] G. P. Agrawal, Applications of Nonlinear Fiber Optics (2001).
[31] T. R. Zhang, H. R. Choo, and M. C. Downer, *Appl. Opt.* **29**, 3927 (1990).
[32] B. Willenberg, F. Brunner, C. R. Phillips, and U. Keller, *Optica* **7**, 485 (2020).
[33] R. J. Essiambre, M. A. Mestre, R. Ryf, A. H. Gnauck, R. W. Tkach, A. R. Chraplyvy, Y. Sun, X. Jiang, and R. Lingle, *IEEE Photonics Technol. Lett.* **25**, 535–538 (2013).
[34] B. A. Malomed, A. Mostofi, and P. L. Chu, *J. Opt. Soc. Am. B* **17**, 507 (2000).
[35] Y. Dong, D. Wang, Y. Wang, and J. Ding, *Phys. Lett. Sect. A Gen. At. Solid State Phys.* **382**, 2006–2012 (2018).
[36] T. L. Wu, and C. H. Chao, *IEEE Photonics Technol. Lett.* **17**, 67–69 (2005).
[37] S. Lee, W. Ha, J. Park, S. Kim, and K. Oh, *Opt. Commun.* **285**, 4082–4087 (2012).
[38] K. M. Hilligsøe, T. V. Andersen, H. N. Paulsen, C. K. Nielsen, K. Mølmer, S. Keiding, K. P. Hansen, R. E. Kristiansen, and J. J. Larsen, *OSA Trends Opt. Photonics Ser.* **96 A**, 1567–1568 (2004).
[39] P. S. Westbrook, J. W. Nicholson, K. S. Feder, Y. Li, and T. Brown, *Appl. Phys. Lett.* **85**, 4600–4602 (2004).
[40] R. Sollapur, D. Kartashov, M. Zürch, A. Hoffmann, T. Grigorova, G. Sauer, A. Hartung, A. Schwuchow, J. Bierlich, J. Kobelke, M. Chemnitz, M. A. Schmidt, and C. Spielmann, *Light Sci. Appl.* **6**, e17124 (2017).
[41] L. B. Michael, M. Ghavami, and R. Kohno, in: 2002 IEEE Conf. Ultra Wideband Syst. Technol. (IEEE Cat. No.02EX580): (2002), pp. 47–51.
[42] H. Zia, *Commun. Nonlinear Sci. Numer. Simul.* **54**, (2018).
[43] G. P. Agrawal, P. L. Baldeck, and R. R. Alfano, *Opt. Lett.* **14**, 137–139 (1989).
[44] P. Tournois, *Opt. Commun.* **140**, 245–249 (1997).
[45] F. Verluise, V. Laude, Z. Cheng, C. Spielmann, and P. Tournois, *Opt. Lett.* **25**, 575 (2000).
[46] (n.d.).

# Supporting Information

**Maintaining Constant Pulse-duration in Highly Dispersive Media using Nonlinear Potentials**

*Haider Zia \**

Haider Zia
University of Twente, Department Science & Technology, Laser Physics and Nonlinear Optics Group, MESA+ Research Institute for Nanotechnology, Enschede 7500 AE, The Netherlands
E-mail: h.zia@utwente.nl

## 1. Signal with Control of section 3.2 at 1 Meter

The position dependent shift in wavelength of the signal pulse, as discussed in Section 3.2, is shown as a global delay in the co-moving frame of reference time window of the simulation.

Likewise, the dispersive +SPM effects of the control pulse results in Airy like side-lobs and a narrowing of the pulse duration (non-linear compression) of the signal pulse, through cross-phase modulation. As indicated in Figure S1 below.

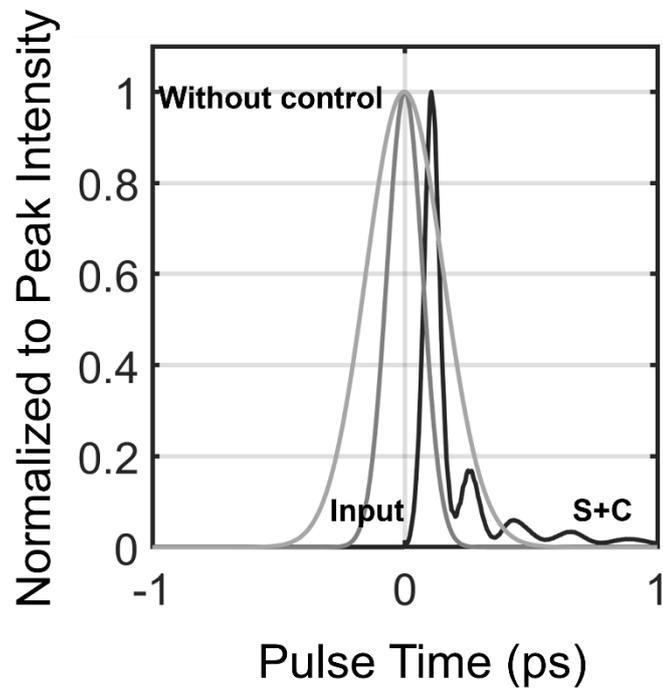

Figure S1: Normalized to peak intensity of the 200 fs Gaussian signal pulse propagated across 1 m of the fiber compared to the input profile and the case without control. Here, the control pulse has full dispersive effects.

## 2. $p$ – Values for Control Pulses

**Table S1.** $p$ – Values for control pulses used for signal pulses illustrated in Figure 1 and Figure 2.

| Figure label of illustrated signal pulse | Approx. P Value (1/m) |
|---|---|
| 1d | 0 |
| 1e | 1.2E3 |
| 1f | 9.8E3 |
| 2b | 0 |
| 2c | 2.0 |
| 2d | 0 |

a)

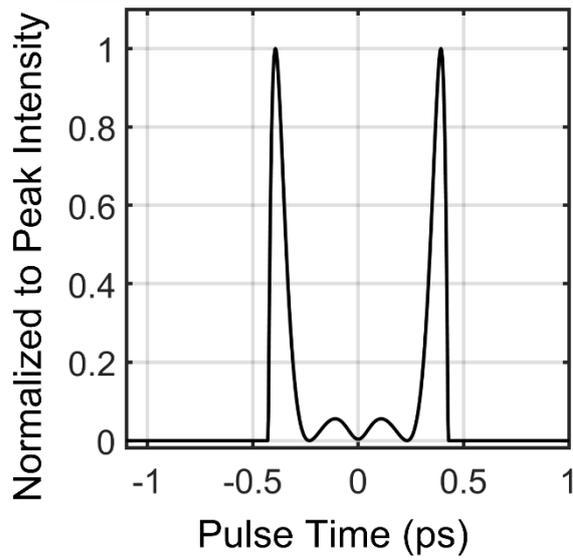

b)

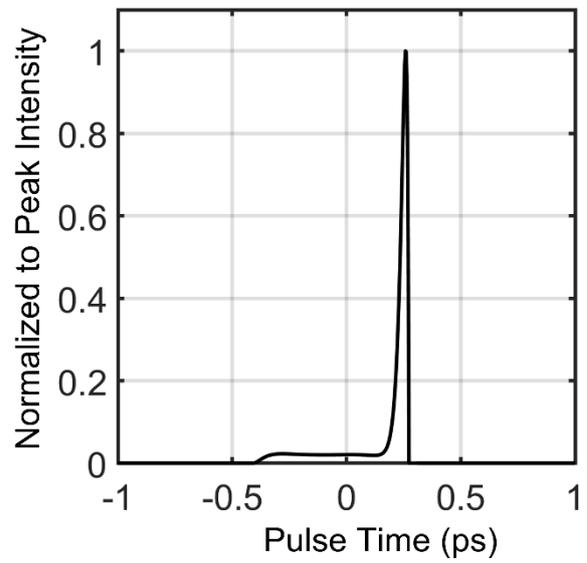

c)

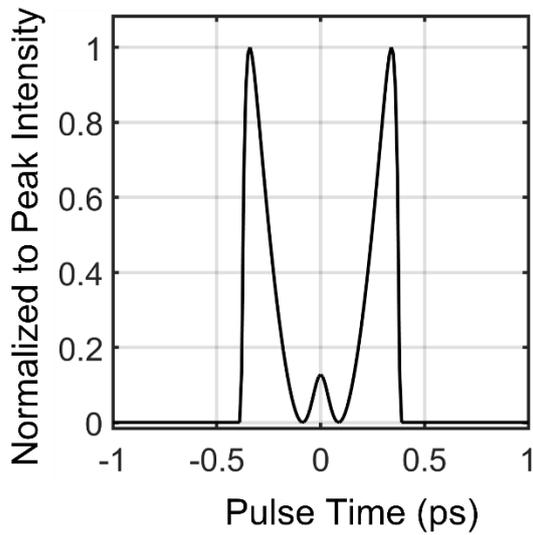

**Figure S2:** a) Normalized intensity of control pulse, plotted against envelope time in picoseconds, for signal pulse of Figure 1e of section 3. b) Normalized intensity of control pulse,

plotted against envelope time in picoseconds, for signal pulse of Figure 1f of section 3. c) Normalized intensity of control pulse, plotted against envelope time in picoseconds, for signal pulse of Figure 2c of section 3.